\documentclass[
    ,final            % use final for the camera ready runs
    ,sort&compress    % add further options here if necessary
  ]
  {aipproc}

\layoutstyle{6x9}
   \usepackage{graphicx}
   \usepackage{epstopdf}
  \usepackage{amssymb}

\def\teq#1{$\, #1\,$}                         % text equation
\def\aa{{Astron. Astrophys.}}

\def\apj{ApJ}

\def\app{Astroparticle Phys.}                   % DO NOT DELETE
\def\apss{Astr. Space Sci.}                     % DO NOT DELETE
\def\asr{Adv. Space Res.}                       % DO NOT DELETE
\def\jgr{J. Geophys. Res.}
\def\mnras{{M.N.R.A.S.}}
\def\prl{Phys. Rev. Lett.}                      % DO NOT DELETE
\def\prd{Phys. Rev. D}                          % DO NOT DELETE
\def\ssr{Space Sci. Rev.}                       % DO NOT DELETE
\newcommand{\vol}[2]{$\,$\rm #1\rm , #2.}           
             \font\sevenrm=cmr7

          \font\sixrm=cmr6       

\def\machson{{M}_{\hbox{\sixrm S}}}
\def\machalf{{M}_{\hbox{\sixrm A}}}

\def\gamsk{\Gamma_1}
\def\erg{\varepsilon_\gamma}
\def\thetascatt{\theta_{\hbox{\sevenrm scatt}}}
\def\thetaBone{\Theta_{\hbox{\sevenrm Bf1}}}

\def\thetaBsone{\Theta_{\hbox{\sevenrm Bs1}}}
\def\betaHTone{\beta_{\hbox{\sevenrm 1HT}}}
\def\uHTone{u_{\hbox{\sevenrm 1HT}}}
\def\muHT{\mu_{\hbox{\sevenrm HT}}}

\def\pF{{p}_{\hbox{\sixrm F}}}
\def\pS{{p}_{\hbox{\sixrm S}}}
\def\dover#1#2{\hbox{${{\displaystyle#1 \vphantom{(} }\over{
   \displaystyle #2 \vphantom{(} }}$}}

\begin{document}
\begin{flushright}
\phantom{p}
\vspace{-60pt}
     To appear in Proc. of the {\it 8th International Astrophysics Conference}\\ 
     ``Shock Waves in Space and Astrophysical Environments,'' (2010),\\
     eds. X. Ao, R. Burrows \& G.~P. Zank (AIP Conf. Proc. 1183, New York).
\end{flushright} 

\title{Particle Acceleration at Relativistic Shocks in Extragalactic Systems}

\author{Matthew G. Baring$^1$ \& Errol J. Summerlin$^{1,2}$}{
    address={$^1$ Department of Physics and Astronomy, MS-108,
                      Rice University, P. O. Box 1892, \\
                      Houston, TX 77251-1892, USA \ \ {\rm Email: baring@rice.edu}\\
                      $^2$ Heliospheric Physics Laboratory, Code 672,
                      NASA's Goddard Space Flight Center,\\
                      Greenbelt, MD 20770, USA \ \  {\rm Email: errol.summerlin@nasa.gov}}
}



\begin{abstract}
Diffusive shock acceleration (DSA) at relativistic shocks is expected to
be an important acceleration mechanism in a variety of astrophysical
objects including extragalactic jets in active galactic nuclei and gamma
ray bursts.   These sources remain strong and interesting candidate
sites for the generation of ultra-high energy cosmic rays. In
this paper, key predictions of DSA at relativistic shocks that are
salient to the issue of cosmic ray ion and electron production are outlined. 
Results from a Monte Carlo simulation of such diffusive acceleration in
test-particle, relativistic, oblique, MHD shocks are presented.
Simulation output is described for both large angle and small angle
scattering scenarios, and a variety of shock obliquities including
superluminal regimes when the de Hoffman-Teller frame does not exist.
The distribution function power-law indices compare favorably with
results from other techniques.  They are found to depend sensitively on the 
mean magnetic field orientation in the shock, and the nature of MHD 
turbulence that propagates along fields in shock environs.
An interesting regime of flat spectrum generation is addressed, providing
evidence for its origin being due to shock drift acceleration.
The impact of these theoretical results on gamma-ray burst and blazar
science is outlined. Specifically, {\it Fermi} gamma-ray
observations of these cosmic sources are already providing significant
constraints on important environmental quantities for relativistic shocks,
namely the frequency of scattering and the level of field turbulence.
\end{abstract}

\maketitle

\section{Introduction}
\label{sec:Introduction}
There is bountiful evidence for efficient particle acceleration at
collisionless shocks in the universe.  The heliosphere, with its
planetary bow shocks and traveling interplanetary shocks, has provided
interesting and useful test cases for shock acceleration theories.  In
the remote regions of the universe, supersonic jets from active galactic
nuclei (AGNs) and gamma-ray bursts (GRBs) have offered fascinating
windows into an energetic part of the cosmos, where non-thermal radio
waves, X-rays and gamma-rays abound.  Fully understanding these sources
mandates knowledge of how the particles that generate their light
emission are energized.  The foremost paradigms invoke acceleration at
relativistic shocks.  Radio imaging reveals very structured and
time-variable jets in AGNs, rapid X-ray and gamma-ray variability in
both blazar AGNs and GRBs suggest compact emission regions with
relativistic bulk motions.  Accordingly, comprehending the relationship
between shock acceleration predictions and observations of these sources
offers the key to elucidating the understanding of their environs; this
constitutes the focus of this paper.

To effect this, an investigation of the features of diffusive shock
acceleration is presented, using results from a test particle Monte
Carlo simulation.  Then the paper addresses probes of the acceleration
theory parameter space imposed by extant GRB and blazar observations in
high energy gamma-rays.  The Monte Carlo approach
\cite{EJR90,ED04,NO04,SBS07} is one of several major techniques devised
to model particle acceleration at relativistic shocks; others include
semi-analytic solutions of the diffusion-convection equation
\cite{KS87,KH89,Kirk00}, and particle-in-cell (PIC) full plasma
simulations \cite{Hoshino92,Nishikawa05,Medvedev05,Spitkovsky08}. Each
has its merits and limitations.  Tractability of the analytic approaches
generally restricts their solutions to power-law regimes for the phase
space distributions \teq{f(\hbox{\bf p})}.  PIC codes are rich in their
information on shock-layer electrodynamics and turbulence.  However, to
interface with astrophysical spectral data, a broad dynamic range in
momenta is desirable, and this is the natural niche of Monte Carlo
simulation techniques. A core property of acceleration at the
relativistic shocks is that the distribution functions \teq{f(\hbox{\bf
p})} are inherently anisotropic. This renders the power-law indices and
other distribution characteristics sensitive to directional influences,
such as the magnetic field orientation with respect to the shock normal,
and the nature of MHD turbulence that often propagates along the field
lines.  These connections between observables and physical parameters of
the shock environs are studied in some depth in this exposition.

\section{Diffusive Acceleration at Relativistic Shocks}
 \label{sec:DSA}
\begin{figure}
 \centerline{
 \hskip -1.5truecm
  \includegraphics[width=.52\textwidth]{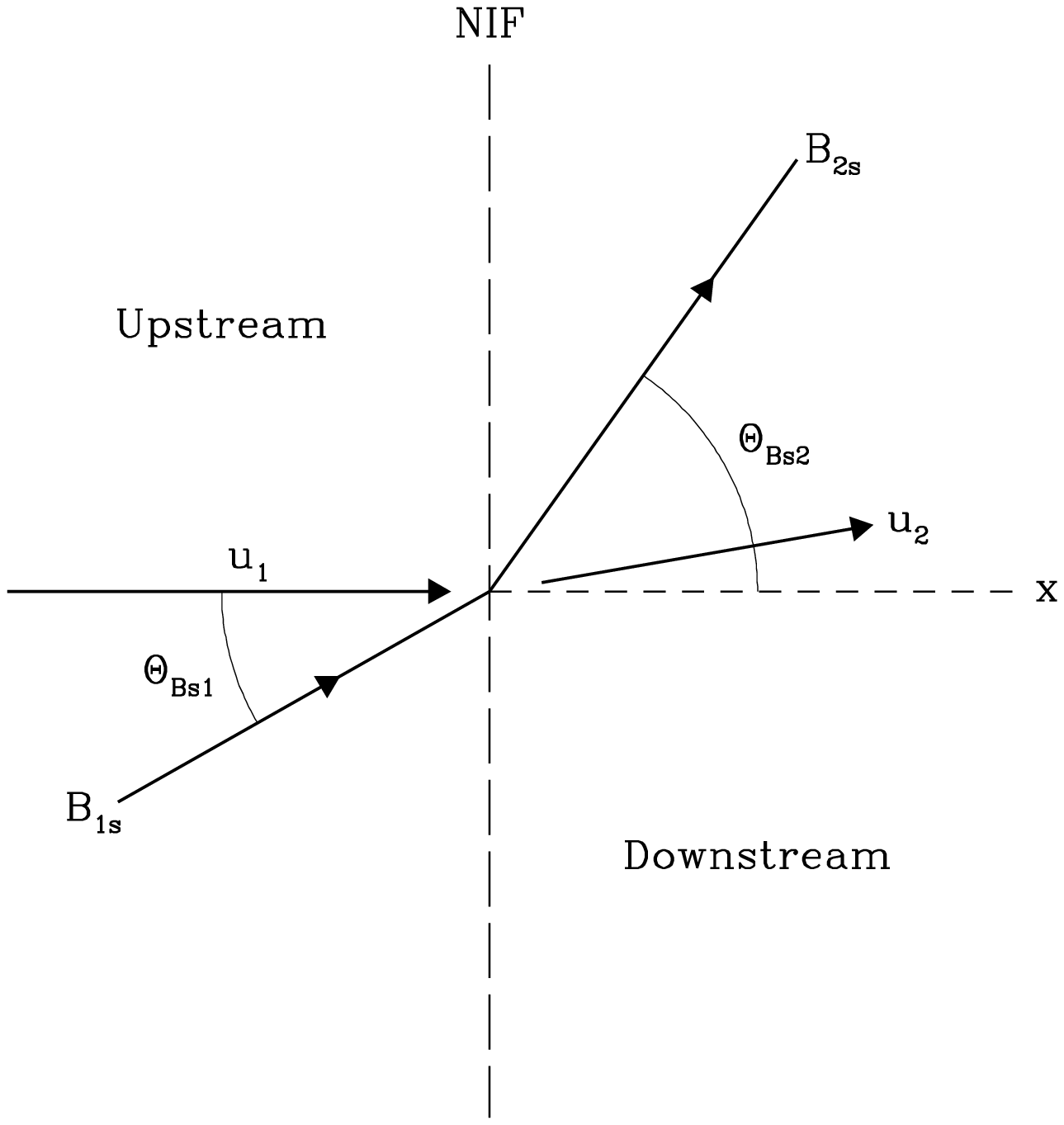}
   \hskip -0.8truecm
    \includegraphics[width=.52\textwidth]{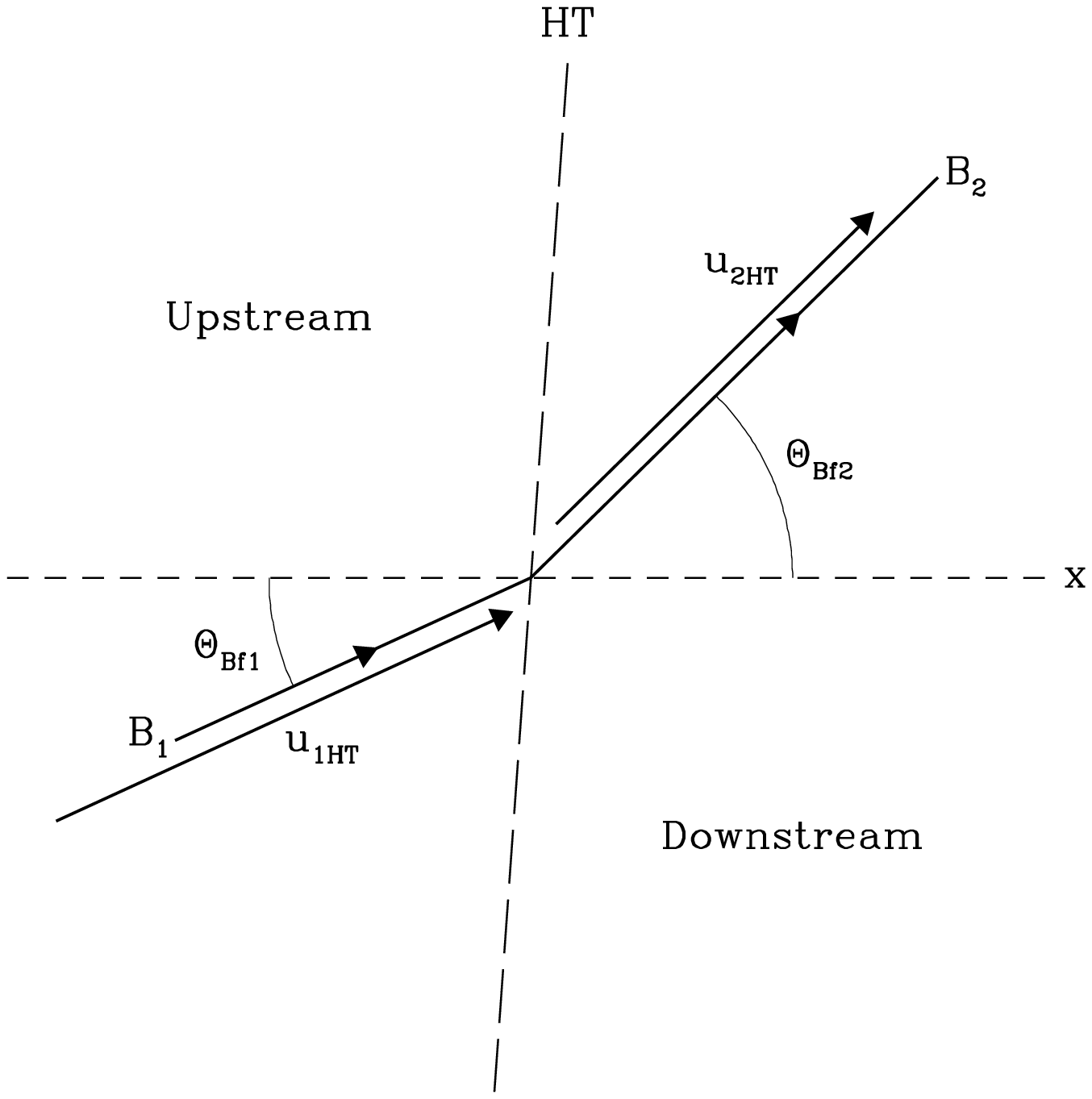}
  }
\caption{The geometry in the normal incidence (NIF; left panel) and 
de-Hoffmann Teller (HT; right panel) shock rest frames.  Upstream 
flow speeds in the two reference frames are related by 
\teq{\uHTone\equiv \betaHTone c= u_1/\cos\thetaBone}.
Upstream and downstream quantities are denoted by subscripts \teq{1} and \teq{2},
respectively.  In general, the NIF field angle \teq{\thetaBsone} differs from the
fluid frame/HT frame value \teq{\thetaBone}.  Also, the shock plane in the HT frame
is rotated from that in the NIF due to relativistic aberration effects.}
 \label{fig:shock_geometry}
\end{figure}

The exploration of the properties of diffusive shock acceleration that
is presented here employs the well-known kinematic Monte Carlo technique
of Ellison and Jones that has been employed extensively in supernova
remnant and heliospheric contexts, and is described in detail in
numerous papers \cite{EJR90,JE91,EBJ95,ED04,SB06,SB09}. Particles are
injected upstream and allowed to convect into the shock, meanwhile
diffusing in space so as to effect multiple shock crossings, and thereby
gain energy through the shock drift and Fermi processes.  In general,
the upstream fluid frame magnetic field is inclined at an angle
\teq{\thetaBone} to the shock normal.  The particles gyrate in laminar
electromagnetic fields, with their trajectories being obtained by
solving the Lorentz force equation in the normal incidence shock rest
frame (NIF), where the upstream flow is incident along the shock normal
($x$-direction: see the left hand panel of Fig.~\ref{fig:shock_geometry}
for the NIF shock geometry).  In this frame, when \teq{\thetaBone > 0},
there is a {\bf u $\times$ B} drift electric field in addition to the
magnetic field. The effects of Alfv\'{e}n wave and other hydromagnetic
magnetic turbulence on particle propagation are modeled by
phenomenologically scattering these ions elastically in the rest frame
of the local fluid flow. This is generally applicable for high Alf\'enic
Mach number shocks.  The simulation outputs particle fluxes and momentum
and angular distributions, usually in the NIF, at any location upstream
or downstream of the shock.

The simulation can routinely model diffusion incurred in particle
interactions with MHD turbulence using either large-angle or small-angle
scattering. At every scattering, the direction of the particle's
momentum vector {\bf p} is deflected in the local fluid frame to a new
value $\hbox{\bf p}'$.  The scattering angle \teq{\delta\theta =
\cos^{-1}(\hbox{\bf p}.\hbox{\bf p}' /\vert\hbox{p}\vert\,\vert\hbox{p}'\vert )} 
is uniformly sampled within a solid angle \teq{\delta\Omega} up to a
maximum deflection angle \teq{\thetascatt} (see Figs.~2 and~3 of
\cite{ED04} for the scattering geometry). The time, \teq{\delta t_f},
between scatterings in this frame is coupled \cite{EJR90} to the mean
free path, \teq{\lambda}, and \teq{\thetascatt}, via \teq{\delta
t_f\approx \lambda\thetascatt^{2}/(6v)} for particles of speed \teq{v}.
The resulting effect is that the gyrocenter of a particle with
gyroradius \teq{r_g} is shifted randomly by a distance of the order of
\teq{\thetascatt r_g} in the plane orthogonal to the local field. 
Accordingly, cross-field diffusion emerges naturally from the
simulation.  For large angle scattering (LAS, defined more precisely
below), the scattering solid angle is \teq{\thetascatt\lesssim \pi}
steradians, and the transport is governed by kinetic theory 
\cite{FJO74,EBJ95}, where the ratio of the spatial diffusion
coefficients parallel (\teq{\kappa_\parallel =\lambda v/3}) and
perpendicular (\teq{\kappa_\perp}) to the mean magnetic field is given
by \teq{\kappa_\perp /\kappa_\parallel = 1/(1+\eta^2)}.  Here, the
parameter \teq{\eta =\lambda/r_g} is the ratio of a particle's mean free
path \teq{\lambda} to its gyroradius \teq{r_g}.  As is often implemented
for simplicity, in this work \teq{\lambda} is assumed to be proportional
to the particle momentum \teq{p}, so that \teq{\eta} is independent of
\teq{p}. This can be adjusted to
accommodate other scattering laws as desired.  Clearly, \teq{\eta}
controls the amount of cross-field diffusion, and is a measure of the
level of turbulence present in the system, i.e. is an indicator of
\teq{\langle \delta B/B\rangle}.  The Bohm limit of quasi-isotropic
diffusion is realized when \teq{\eta\sim 1} and \teq{\langle \delta
B/B\rangle\sim 1}.   This phenomenological description of diffusion in
Monte Carlo techniques is most appropriate at high energies, and omits
the details of microphysics present in plasma simulations such as PIC
codes.  In the injection domain at slightly suprathermal energies, the
influences of complex turbulent and coherent electrodynamic effects
become important, and will substantially modify the picture from pure
diffusion.

%% FIGURE 1 GOES HERE

It will become clear below that in oblique relativistic shocks, the
diffusive transport of particles across the field, and hence across the
shock, becomes critical to their retention in the acceleration process.
Accordingly, for such systems, the interplay between the field angle
\teq{\thetaBone} and the value of \teq{\lambda /r_g} controls the
spectral index of the particle distribution \cite{ED04,Baring04}, a
feature that is central to the interpretation of astrophysical source
spectra.  The phase space for acceleration theory results is now
explored, before addressing their relevance to two classes of cosmic
sources: gamma-ray bursts and blazars.

\section{Acceleration Signatures at Relativistic Shocks}
 \label{sec:accel_prop}
Diffusive particle acceleration at relativistic shocks naturally divides
into two regimes: mildly-relativistic (\teq{u_{1x}\lesssim c}) and
ultra-relativistic (\teq{u_{1x}\approx c}) systems. Hereafter,
subscripts \teq{1} denote upstream quantities. Neither regime is
accessible to diagnostics via {\it in situ} particle measurements since
the principal astrophysical sources possessing relativistic shocks
exhibiting signatures of acceleration, namely pulsar wind termination
shocks, jets in active galactic nuclei (blazars) and gamma-ray bursts,
are so distant. An important property of diffusive acceleration at
relativistic shocks that distinguishes them from their non-relativistic
counterparts is their intrinsic anisotropy. This is driven by the
powerful convective influence that enables efficient loss of particles
away and downstream from the shock.  The immediate consequence of this
loss is a general difficulty in generating flat distributions of
shock-accelerated particles, particularly for so-called superluminal
(\teq{\beta_{1x}/\cos\thetaBone > 1}) relativistic discontinuities.
These losses are muted somewhat in mildly-relativistic shocks, which are
naturally expected in GRBs and perhaps also in blazars. Moreover, the
resulting distributions \teq{dN/dp \propto p^2 f(\hbox{\bf p})} are more
or less commensurate with those required to explain the radiation
observations from these sources. Accordingly, this paper will study
\teq{\gamsk = (1-u_{1x}^2/c^2)^{-1/2} \sim 1 - 5} shocks, focusing on
the power-law tails \teq{dN/dp\propto p^{-\sigma}} realized at high
momenta, where all memory of the injection momentum has been destroyed
by diffusion. The array of possible distribution indices \teq{\sigma} is
highlighted, spawned by the sensitivity of both the energization in, and
escape from, the shock layer, to (i) the size of the momentum deflection
angle \teq{\thetascatt}, (ii) the frequency or relative mean free path
\teq{\lambda/r_g} of scatterings, and (iii) the upstream field obliquity
\teq{\thetaBone}, a quantity connected to the global MHD structure of
the shock.

The first effect, the dependence of \teq{\sigma} on \teq{\thetascatt},
has been documented in several recent papers \cite{Baring04,ED04,SBS07},
and will just be summarized here.  When the diffusion in the shock layer
samples large field fluctuations \teq{\delta B/B\sim 1} (seen, for
example, in PIC simulations of relativistic shocks driven by the Weibel
instability \cite{Hoshino92,Nishikawa05,Medvedev05,Spitkovsky08}), it
corresponds to large momentum deflections, delineating the regime of
large angle scattering (LAS) with \teq{4/\Gamma_1\lesssim
\thetascatt\lesssim\pi}), where \teq{\Gamma_1} is the upstream flow's
incoming Lorentz factor. This regime was first explored for
\teq{\Gamma_1\lesssim 5}  by Ellison et al. \cite{EJR90}. Such large
deflections produce huge gains in particle energy, of the order of
\teq{\Gamma_1^2}, in successive shock crossings. These gains are
kinematic in origin, and are akin to those in inverse Compton
scattering.  The result is an acceleration distribution \teq{dN/dp} that is 
highly structured and much flatter on average \cite{EJR90} than \teq{p^{-2}} 
for strong, parallel (\teq{\thetaBone =0^\circ}), shocks, i.e. those with high
sonic and Alf\'enic Mach numbers. The bumpy structure is kinematic in
origin, corresponding to sequential shock transits \cite{Baring99}, and
becomes more pronounced \cite{Baring04,ED04,SBS07,B09} for large
\teq{\Gamma_1}.  For ultra-relativistic shocks, information of the
injection momentum scale becomes insignificant when \teq{p\gg mc}, and
the bumps asymptotically relax to form a power-law distribution
\teq{dN/dp\propto p^{-\sigma}}, with an index in the range of
\teq{\sigma\sim 1.6} \cite{SBS07}.  From the plasma physics perspective,
magnetic turbulence in relativistic shocks could easily be sufficient 
to effect scatterings on intermediate to large angular scales
\teq{\thetascatt\gtrsim 1/\Gamma_1}, a proposition that becomes more
enticing for ultra-relativistic shocks.

The principal focus in this paper is on particle distributions for
\teq{\thetascatt\lesssim 1/\Gamma_1} regimes, which are much smoother in
appearance, and often necessarily steeper, at least for superluminal
regimes. The property of distribution smoothness meshes more easily with
radiation spectral observations of extragalactic astrophysical sources,
thereby motivating exploration of this portion of phase space. The
kinematic energy gains in shock crossings are lowered considerably
\cite{Baring99} when \teq{\thetascatt} drops below this ``Lorentz cone''
angle \teq{1/\gamsk}. Accordingly, the character of \teq{dN/dp} and the
particle anisotropy at the shock dichotomize, partitioned by the
\teq{\thetascatt\sim 1/\gamsk} boundary. Intermediate scattering angles
\teq{\thetascatt\sim 1/\Gamma_1} generate smooth distributions
\cite{SBS07,B09}, much like those for small angle scattering (SAS, often
called pitch angle diffusion, PAD). The SAS regime has spawned the often
cited asymptotic, ultrarelativistic index of \teq{\sigma =2.23} for
\teq{dN/dp\propto p^{-\sigma}} \cite{Kirk00}, first noticed in Monte
Carlo simulations \cite{BO98}.  This special result, applicable in both
shock rest and fluid frames for momenta \teq{p\gg \Gamma_1mc}, is
realized only for parallel shocks with \teq{\thetaBone =0^{\circ}} in
the limit of \teq{\thetascatt\ll 1/\Gamma_1}, where the particle
momentum is stochastically deflected on arbitrarily small angular (and
therefore temporal) scales.  In such cases, particles diffuse in the
region upstream of the shock only until their velocity's angle to the
shock normal exceeds around \teq{1/\Gamma_1}, after which they are
rapidly swept downstream of the shock. The lower kinematic energy gains
in shock transits dominate higher shock-layer retention rates, and
guarantee a steeper distribution under SAS conditions for
\teq{\thetaBone =0^{\circ}} shocks; the monotonic steepening and loss of
structure as \teq{\thetascatt} declines is exhibited in \cite{SBS07} for
ultra-relativistic shocks, and \cite{B09} for the mildly-relativistic domain.

\begin{figure}
 \centerline{
  \includegraphics[width=.51\textwidth]{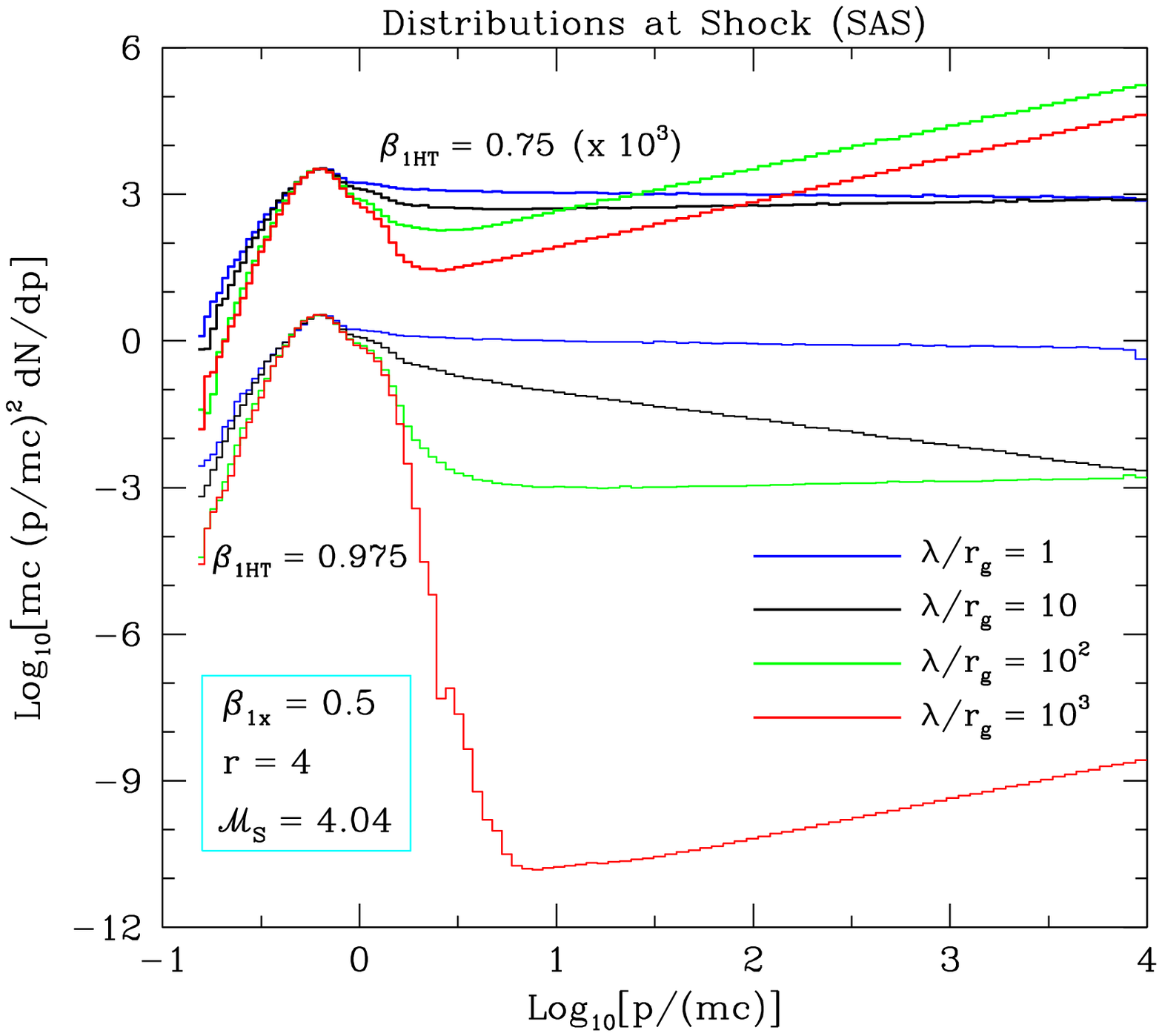}
   \hskip -0.1truecm
    \includegraphics[width=.51\textwidth]{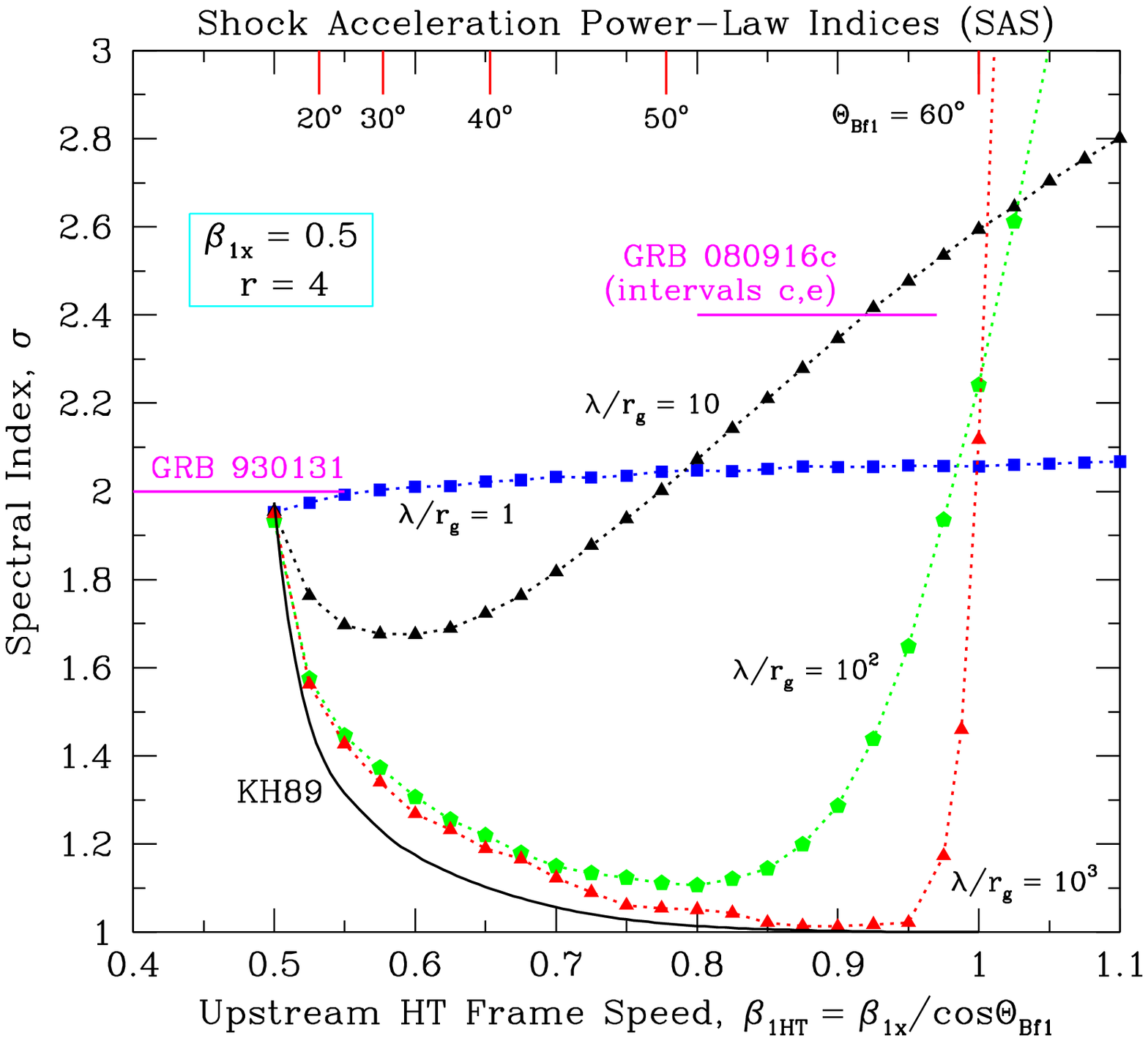}
  }
  \caption{{\it Left panel:}
Particle distribution functions  \teq{dN/dp} (scaled by \teq{p^2}) from
mildly-relativistic sub-luminal shocks
(\teq{\Gamma_{1x}\beta_{1x}=0.577}, i.e. \teq{\beta_{1x}=u_{1x}/c=0.5})
of upstream-to-downstream  velocity compression ratio
\teq{r=u_{1x}/u_{2x}\approx 4}.  Simulation results are depicted for two
de Hoffman-Teller frame upstream flow speeds \teq{\beta_{\hbox{\sevenrm
1HT}} = \beta_{1x}/\cos\thetaBone}. These are in distinct groups of
four: \teq{\beta_{\hbox{\sevenrm 1HT}} = 0.75} (upstream magnetic field
obliquity \teq{\thetaBone=48.2^{\circ}}, multiplied by \teq{10^3}) for
the upper four histograms, and \teq{\beta_{\hbox{\sevenrm 1HT}} = 0.975}
(i.e., \teq{\thetaBone=59.1^{\circ}}) for the lower four histograms.
Scattering off hydromagnetic turbulence was modeled by randomly
deflecting particle momenta by an angle within a cone, of half-angle
\teq{\thetascatt},  whose axis coincides  with the particle momentum
prior to scattering; four different ratios of the diffusive mean free
path \teq{\lambda} to the gyroradius \teq{r_g} were adopted for each
\teq{\thetaBone}.  All results were for small angle scattering (SAS),
when \teq{\thetascatt\lesssim 1/\Gamma_1} and the distributions become
independent of the choice of \teq{\thetascatt}.
\newline
{\it Right panel:}
Power-law indices \teq{\sigma} for simulation runs in the limit of small
angle scattering (pitch angle diffusion), for mildly-relativistic shocks
of upstream flow speed \teq{\beta_{1x}\equiv u_{1x}/c =0.5}, and an MHD
velocity compression ratio \teq{r=4}. The indices are displayed as
functions of the effective de Hoffman-Teller frame upstream flow speed
\teq{\beta_{\hbox{\sevenrm 1HT}} = \beta_{1x}/\cos\thetaBone}, with
select values of the fluid frame field obliquity \teq{\thetaBone} marked
at the top of the panel. The displayed simulation index results were
obtained for different diffusive mean free paths \teq{\lambda} parallel
to the mean field direction, namely \teq{\lambda/r_g=1} (squares),
\teq{\lambda/r_g=10} (triangles), \teq{\lambda/r_g=10^2} (pentagons),
and \teq{\lambda/r_g=10^3} (triangles), as labelled.  The lightweight
curve at the bottom labelled KH89 defines the semi-analytic result from
Kirk \& Heavens' \cite{KH89} solution to the diffusion-convection
equation, corresponding to \teq{\lambda/r_g\to\infty}. The short
heavyweight lines labelled GRB 930131 (EGRET detection) and GRB 080916c
({\it Fermi} detection) indicate the approximate spectral index
\teq{\sigma} that is appropriate for these gamma-ray bursts, if a cooled
synchrotron emission scenario is operable.
}
 \label{fig:sas_spec_index}
\end{figure}

Now the focus turns to the influence the effective frequency
\teq{\lambda/r_g} of scatterings, and the upstream field obliquity
\teq{\thetaBone} have on the accelerated population. Representative
particle (electrons or ion) differential distributions \teq{dN/dp} that
result from the simulation of diffusive acceleration at
mildly-relativistic shocks of speed \teq{\beta_{1x}=0.5} are depicted in
the {\it left panel} of Figure~\ref{fig:sas_spec_index} (see
\cite{ED04,SBS07} for \teq{\Gamma_1\gg 1} simulation results).  Here,
the subscript \teq{x} denotes components along the shock normal.  These
distributions were generated for \teq{\thetascatt \lesssim 10^\circ},
i.e. in the SAS regime, for low magnetic fields corresponding to
Alfv\'enic Mach numbers \teq{\machalf\gg 1}, and in the NIF frame.
Results are displayed for two different upstream fluid frame field
obliquities, namely \teq{\thetaBone=48.2^{\circ}} and
\teq{\thetaBone=59.1^{\circ}}, with corresponding de Hoffman-Teller (HT
\cite{dHT50}) frame dimensionless speeds of \teq{\betaHTone
=\beta_{1x}/\cos\thetaBone = 0.75} and \teq{0.975}, respectively. {\it
Subluminal} shocks are those where the HT flow speed \teq{\betaHTone}
corresponds to a physical speed, less than unity, i.e. the upstream
field obliquity satisfies \teq{\cos\thetaBone < \beta_{1x}}.  When
\teq{\betaHTone > 1}, the de Hoffman-Teller frame does not exist, and
the shock is said to be {\it superluminal}. See the right hand panel of
Fig.~\ref{fig:shock_geometry} for the HT frame shock geometry. The
distributions clearly exhibit an array of indices \teq{\sigma},
including very flat power-laws, that are not monotonic functions of
either the field obliquity \teq{\thetaBone} nor the key diffusion
parameter \teq{\eta =\lambda /r_g}.  These properties are illustrated in
the {\it right panel} of Fig.~\ref{fig:sas_spec_index}, where it is also
evident that the distributions are generally steeper in superluminal
shocks \cite{ED04} with \teq{\lambda/r_g\gtrsim 10}.  The left panel of
Fig.~\ref{fig:sas_spec_index} also emphasizes that the normalization of
the power-laws relative to the low momentum thermal populations is a
strongly-declining function of \teq{\lambda /r_g}.  This is a
consequence of a more prolific convection of suprathermal particles
downstream of the shock that suppresses diffusive injection from thermal
energies into the acceleration process.  Such losses are even more
pronounced when \teq{\lambda /r_g \geq 10^4}, to the point that
acceleration is not statistically discernible for \teq{\betaHTone >
0.98} runs with \teq{10^4} simulated particles. This feature is salient
for the discussion on astrophysical sources below.

% FIGURE 2 GOES HERE

The existence of very flat distributions in the subluminal domain for
very large \teq{\lambda /r_g} is a remarkable feature of
Figure~\ref{fig:sas_spec_index}.  This phenomenon was identified by Kirk
\& Heavens \cite{KH89} in their eigenfunction solution technique
\cite{KS87} for the diffusion-convection equation, which was restricted
to subluminal, oblique shocks.  Results for their  \teq{\beta_{1x}=0.5}
analysis are presented as the solid curve labelled KH89 in the right
panel of the Figure.  Clearly, the Monte Carlo indices closely approach
those of the semi-analytic method of \cite{KH89} for \teq{\lambda
/r_g=10^3}, an agreement that is improved slightly when \teq{\lambda
/r_g} is increased to \teq{10^4}.  Yet there are differences between the
two approaches, and these yield the expected slight discrepancies in
spectral index determination.  The method of \cite{KH89}, being
tantamount to a guiding center technique, employs conservation of the
magnetic moment \teq{p^2 (1-\mu^2)/(2B)} for particle-shock interactions when
determining transmission and reflection probabilities of charges,
naturally differing from gyro-orbit determinations of these
probabilities.  This nuance probably seeds the different values
of \teq{\sigma} derived in the two approaches, which, as expected, are
small when shocks are effectively parallel, and highly oblique. More
details of the comparison of these approaches are discussed in
\cite{SB09}, which provides an extended exposition on the implementation
of the Monte Carlo code, its validation and many of its key acceleration
results. It can also be noted that comparing with the indices derived by
\cite{KH89} was the major motivation behind the artificial choice of the
compression ratio \teq{r=4}, which is somewhat larger than the
Rankine-Hugoniot MHD value for \teq{\beta_{1x}=0.5}, \teq{\machson
=4.04} conditions. Also, a low sonic Mach number \teq{\machson} was
chosen so as to maximize the efficiency of injection from thermal
energies.

\subsection{The Action of Shock Drift Acceleration}
\begin{figure}
 \centerline{
  \includegraphics[width=.51\textwidth]{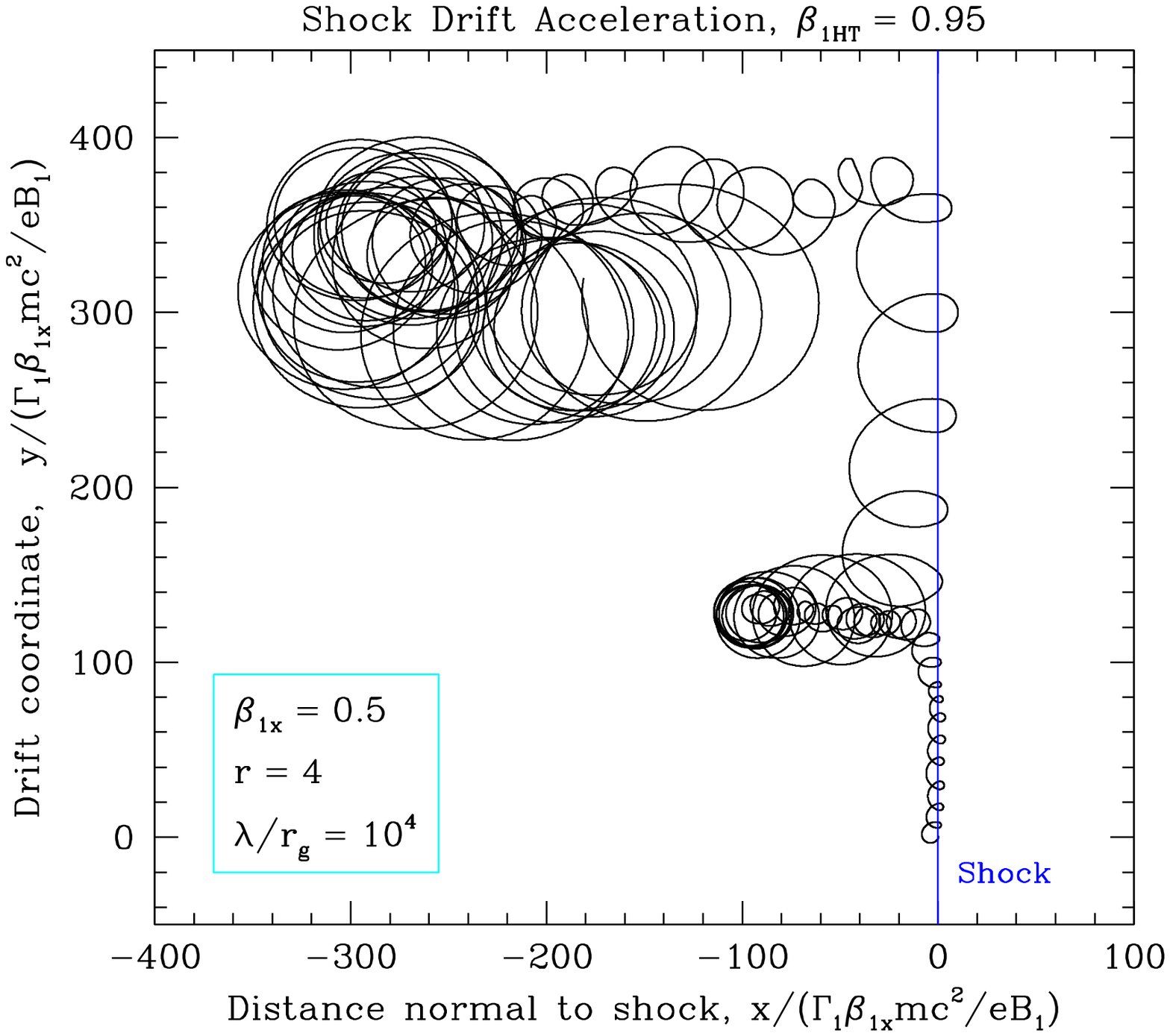}
   \hskip -0.1truecm
    \includegraphics[width=.51\textwidth]{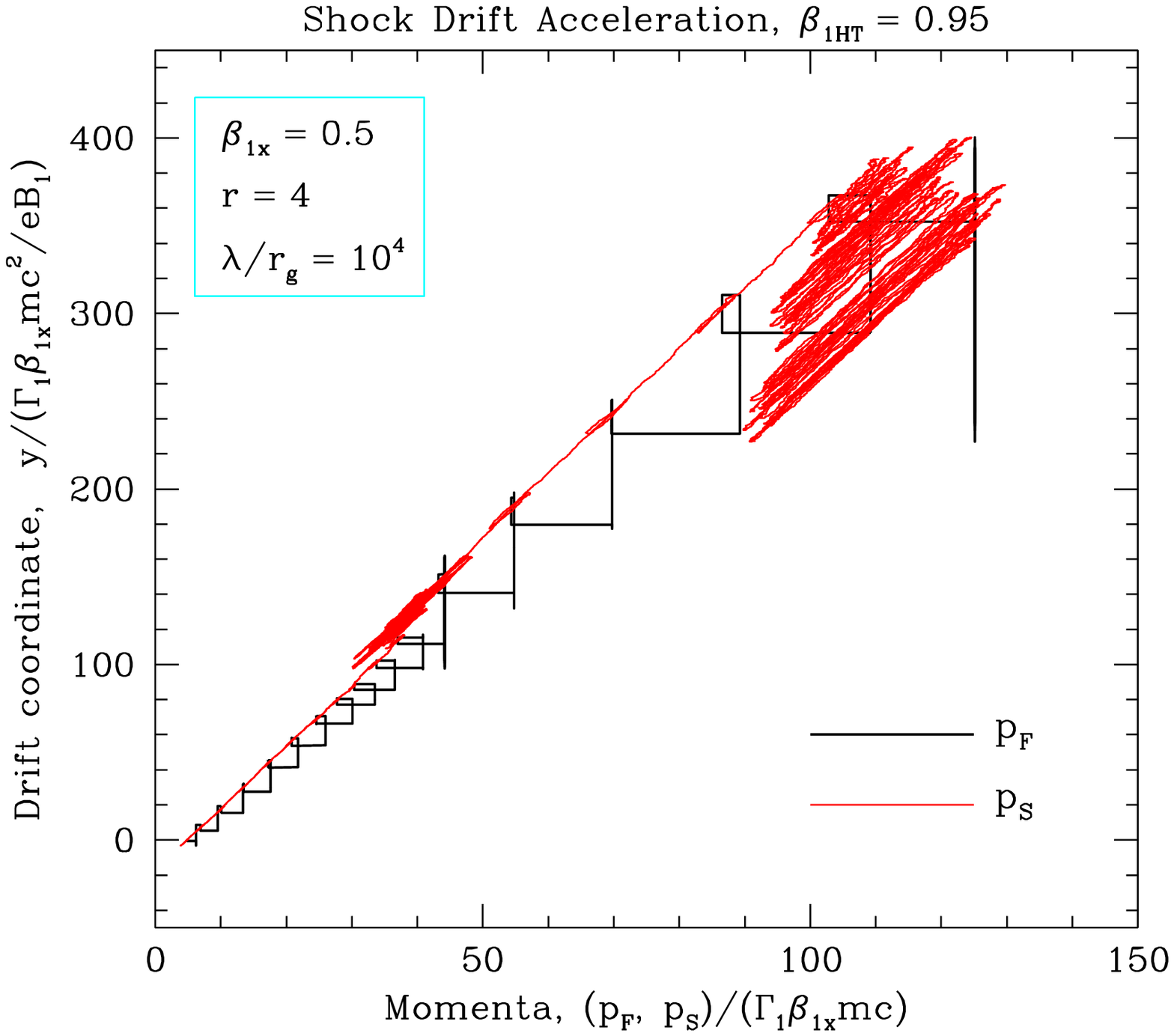}
  }
  \caption{{\it Left panel:}
A portion of a shock-layer trajectory for a selected particle
accelerated from low to high energies, illustrating two key features of
shock drift acceleration: coherent trapping in the shock, and occasional
upstream excursions.  The projection is onto the \teq{x}-\teq{y} plane,
where the {\bf u}\teq{\times}{\bf B} drifts lie in the
\teq{y}-direction.  The shock speeds are as in
Fig.~\ref{fig:sas_spec_index}; the large value of \teq{\lambda/r_g=10^4}
is adopted to generate the requisite conditions for coherent interaction
with the shock. In this projection, energy gains incurred while the
particle's gyrations straddle the shock only become obvious when its
pitch angles are altered in upstream transits. Particles that undergo
these trajectories are rare, but populate the power-law tail.
\newline
{\it Right panel:}
The coupling between the fluid frame (\teq{\pF}) and NIF shock rest
frame (\teq{\pS}) momenta of the particle selected in the left panel, and
the drift coordinate \teq{y}.  The shock frame momentum evinces
oscillatory behavior during upstream excursions, and both \teq{\pF} and
\teq{\pS} display the linear trend with drift \teq{y} that is
characteristic of shock drift acceleration. The fluid frame momentum
exhibits a ``rectangular hysteresis'' while the particle gyrates in the
shock layer.
}
 \label{fig:sas_traj_mom}
\end{figure}

The origin of these flat indices is elucidated by the parameter survey
here, and more specifically by the inclusion of cross field diffusion in
the Monte Carlo simulations.  Diffusive transport of particles
perpendicular to the mean field was omitted in the analysis of
\cite{KH89}, a restriction that is the primary reason for \teq{\sigma}
dropping below two.  This is clearly evident from the right panel of
Figure~\ref{fig:sas_spec_index}, which indicates that \teq{\sigma
\lesssim 1.5} scenarios are realized only for \teq{\lambda /r_g\gtrsim
10^2}, i.e. laminar fields and almost pure gyrational motion.  The primary
origin of the acceleration is then connected to coherence in the shock
layer.  To provide further insight, individual particle trajectories
were tracked in the Monte Carlo runs, and those exhibiting profound
energy gains isolated.  One such example is provided in
Fig.~\ref{fig:sas_traj_mom}, together with its corresponding momentum 
trace (right panel); the particles were injected with
superthermal momenta \teq{p=5mc} to circumvent any injection problems.
Diffusion has minimal impact on the gyrational motion in the left panel.
The trajectories are 2D projections, and the pitch angle is evolving
almost adiabatically in the particle-shock interaction to preserve
gyrational coherence. The momentum histories in the right hand panel
more readily highlight the property that the acceleration is directly
coupled to periods when the particle's gyration straddles the shock.
Moreover, inertial motion in the \teq{y} or {\bf u}$\times${\bf B}
direction accompanies these epochs of energization.  These two
characteristics are the hallmarks of {\it shock drift acceleration} (SDA).

%\phantom{p}
%\vskip -25pt
%

% FIGURE 3 GOES HERE

The acceleration of particles in the drift electric fields associated with 
oblique shock discontinuities has been extensively studied in non-relativistic 
contexts \cite{SvA74,Jok82,PDA82,WAT83,DV86}.  The origin of the effect is 
the net work done on a charge by the Lorentz force in a zone of non-uniform 
magnetic field.  The principal equation governing this is
\begin{equation}
   \hbox{\bf p}.\dover{d\hbox{\bf p}}{dt} \; =\; q\, \hbox{\bf p}.\left\{ \hbox{\bf E} 
                  + \dover{\hbox{\bf v}}{c} \times \hbox{\bf B} \right\}
                  \;\equiv\; q\, \hbox{\bf p}.\hbox{\bf E} \quad ,
 \label{eq:shock_drift}
\end{equation}
where {\bf E} is the {\bf u}$\times${\bf B} drift field that exists in any
oblique shock rest frame other than the HT frame.  In the uniform {\bf B} fields
either upstream or downstream, the energy gains and losses acquired 
during a gyroperiod exactly cancel, so that no net work is done, \teq{dW =0}.
In contrast, when a charge's gyromotion straddles the shock 
discontinuity, the sharp field gradient induces an asymmetry 
in the time spent by the charge either side of the shock, so that 
energy gains and losses do not negate each other.  The compressive 
nature of the field discontinuity biases the net work done to positive 
increments in shock encounters between upstream excursions, and 
it is simply shown \cite{Jok82,WAT83} that \teq{dW = qE_y\, dy}, i.e. 
this energy gain scales linearly with displacement along the 
drift coordinate \teq{y}.  This is the punchline of the right panel
of Fig.~\ref{fig:sas_traj_mom}, where \teq{y} effectively represents 
a time coordinate during shock drift episodes, with \teq{dW/dt = qE_y/c}.  
It is also noteworthy that interspersed between these acceleration periods
are upstream excursions where infrequent scattering slowly tries to
isotropize the pitch angles, a feature clearly identified for the SDA
phenomenon at non-relativistic shocks \cite{DV86}.  Particles that
participate in the SDA initially have HT frame pitch angle
cosines \teq{\muHT} considerably less than unity, so as to satisfy the    
reflection criterion: therefore they are rare in near-luminal shocks, for
which the incident upstream angular distribution is highly-beamed around
\teq{\muHT\sim 1}. During upstream excursions that follow reflection, the
particles tend to get swept back to the shock as soon as they are
deflected and acquire momenta outside their Lorentz cone, i.e. before   
reaching isotropy in the upstream fluid frame.  Therefore, in these
\teq{\lambda /r_g\gg 1} cases, \teq{\muHT} continues to satisfy the 
reflection criterion during these brief upstream epochs.  The repetition 
of SDA and upstream excursions, i.e. trapping in and near the shock, is 
therefore virtually guaranteed once the initial reflection is realized for 
a select particle.

This investigation provides an identification of the significance of
shock drift acceleration in controlling \teq{\sigma} for relativistic
shocks. The reason it couples to unusually flat distributions
(\teq{\sigma < 1.5}) revolves around extremely efficient trapping in the
shock layer, which permits repeated episodes of SDA in select particles
with appropriately-tuned gyrophases at the onset of shock-orbit
interactions. The combination of the rapid energization rate, upstream
hiatuses in SDA, and a slow leakage rate downstream leads naturally
\cite{BS09} to an approximate \teq{dN/dp\propto p^{-1}} distribution. 
But only when \teq{\lambda /r_g\gtrsim 10^3}.  The introduction of
turbulence easily disrupts the coherence and precipitates efficient
convection downstream \cite{BS09}, quenching the effectiveness of SDA. 
The result then is a dominance of shock drift acceleration contributions
by first order Fermi (turbulent) ones when \teq{\lambda/r_g\lesssim 30}.
More importantly, when \teq{\lambda/r_g \gtrsim 10}, an inexorable
sweeping of charges downstream in superluminal shocks overpowers both
acceleration contributions and steepens the spectrum dramatically, as is
evident in the right panel of Fig.~\ref{fig:sas_spec_index}.

\section{Astrophysical Source Context}
 \label{sec:astro}

The shock acceleration theory results presented in the previous Section
can now be interpreted in the light of observations of astrophysical
sources. The first class of germane sources consists of gamma-ray bursts
(GRBs), whose prompt emission is generally observed in the 10 keV - 10
GeV range, with a characteristic spectral break around 200 keV to 1 MeV.
Above this break, the spectrum is generally (but not always) an extended
power-law \teq{dn_{\gamma}/d\erg\propto\erg^{-\alpha_h}}. The focus here
is on the relationship between the high energy spectral index
\teq{\alpha_h} and the underlying particle acceleration conditions. 
This forges a direct connection to data from CGRO's EGRET telescope, and
now to the growing database of {\it Fermi} LAT burst detections.  The
EGRET \teq{\alpha_h} index distribution \cite{Dingus95} is constituted
by a handful of sources with indices scattered in the range
\teq{2\lesssim\alpha_h\lesssim 3.7}, with brighter bursts' indices
concentrated in the range \teq{\alpha_h \lesssim 2.8}, as tabulated in
\cite{Baring06}.  The recent {\it Fermi} detection \cite{Abdo09} of GRB
080916c in both the GBM and LAT instruments offered an index of
\teq{\alpha_h\sim 2.2} at energies above \teq{\sim 2} MeV in its most
luminous epoch, and a steeper spectrum (\teq{\alpha_h\gtrsim 2.5}) at
other times. It is then evident that observationally, shock acceleration
models must accommodate a radiation spectral index in the range
\teq{2\lesssim\alpha_h\lesssim 4} in order to be viable.  Moreover, they
must reasonably account for the spectral variability identified in GRB
080916c, i.e. fluctuating \teq{\alpha_h} values in a given source.

If one presumes that these photon spectra result from synchrotron
emission that rapidly cools the radiating electrons, a popular paradigm
\cite{RM92,Piran99,Meszaros02} for the production of the prompt
emission, then \teq{\alpha_h = (\sigma + 2)/2} for electron acceleration
populations \teq{dN/dp \propto p^{-\sigma}}.  The \teq{\sigma}
corresponding to the values of \teq{\alpha_h=2.0} for the flat-spectrum
EGRET burst GRB 930131 and \teq{\alpha_h=2.2} for GRB 080916c for time
intervals (c,e) \cite{Abdo09} are marked on the right panel of
Fig.~\ref{fig:sas_spec_index}.  It then becomes clear that the emission
in GRB 930131 is consistent with acceleration at subluminal shocks and
relatively near the Bohm diffusion limit.  In contrast, the slightly
steeper GRB 080916c spectrum is better explained by mildly superluminal
shocks in the \teq{60^\circ}-\teq{70^\circ} obliquity range, but only if
the scattering is strong, i.e. \teq{\lambda /r_g\lesssim 3-10}.  If,
instead, synchrotron cooling is inefficient, the photon differential
spectral index given by \teq{\alpha_h = (\sigma + 1)/2}. Then the
injected distribution must have an index \teq{\sigma} higher by unity
than that for cooling models, in order to match the burst observations.
This is a profound difference in that it moves the viable shock
parameter space into the superluminal range, i.e. at higher field
obliquities, and Bohm-limited diffusion is observationally excluded. No
bursts have so far evinced extended power-law spectra flatter than
\teq{\alpha_h\approx 1.95}, absolving the need for acceleration in
shocks with extremely low turbulence, i.e. \teq{\lambda/r_g\gtrsim 10^2}
regimes.  This is fortunate, since, from the left panel of
Fig.~\ref{fig:sas_spec_index}, such shocks are inherently inefficient
accelerators. Moreover, the generation of field turbulence is a natural
part of dissipation in shocks, so that almost laminar fields are not
expected, nor observed in {\it in situ} magnetometer measurements at
heliospheric shocks (e.g. \cite{Balogh95,BOEF97} and references
therein).

The second relevant astrophysical context concerns blazars, the subset
of active galactic nuclei possessing relativistic jets of material
emanating from the supermassive black holes at their centers; these jets
are oriented virtually towards the observer. These have been the
preserve of gamma-ray experiments ever since their discovery by EGRET
\cite{Hart92}, and subsequent observation by ground-based Cherenkov
telescopes at TeV energies \cite{Punch92}.  The TeV-band signals
typically exhibit steep photon spectra (e.g. Mkn 421:
\cite{Krenn02,Ahar03}) that include the absorption due to pair producing
interactions \teq{\gamma\gamma\to e^+e^-} with infra-red and optical
light generated by the intergalactic medium along the line of sight to
the observer.  The correction for this attenuation (so called {\it
de-absorption}), leads to the inference of extremely flat particle
distributions in energetic gamma-ray blazars (see for example
\cite{SBS07}), with indices as low as \teq{\sigma\sim 1.0 - 1.5} in high
redshift sources. To accommodate these observational constraints would
require subluminal shocks with very modest or low turbulence levels. 
Yet, such inferences are based upon measurements in a limited waveband
subject to profound absorption. More improved diagnostics are now
enabled by {\it Fermi} LAT detections of blazars, which extend the
observational window over a much larger energy range, nominally from 100
MeV to over 1 TeV, and most crucially, below the attenuation window.
Accordingly, {\it Fermi} observations can more directly probe the
underlying radiating particle population.  A prime example of this is
the multi-wavelength campaign on the PKS 2155-304 blazar
\cite{Ahar_etal09} in a non-flare state, whose combined {\it Fermi}-{\it
HESS} spectrum from 300 MeV to 3 TeV indicates an unattenuated photon
spectral index of \teq{\alpha_h\sim 1.6}, steepening to
\teq{\alpha_h\sim 2.0} above 1 GeV. For a standard inverse Compton
scattering interpretation with insignificant radiational cooling, this
translates to an electron power-law index in the range \teq{\sigma \sim
2.2\to 3.0}, so that acceleration at mildly superluminal oblique shocks
should provide the best description.

\section{Conclusions}
 \label{sec:conclusion}

This paper has investigated some of the key characteristics of particle
acceleration at relativistic shocks, including the identification of the
role of shock drift acceleration in generating flat distributions in
mildly-relativistic, subluminal shocks.   It has also explored the
connection between the acceleration process and high energy emission in
two classes of astrophysical sources, namely gamma-ray bursts and jets
in blazars. The simulation results presented clearly highlight the
non-universality of the index of energetic, non-thermal electrons and
ions, spawned by the variety of shock obliquities and the character of
hydromagnetic turbulence in their environs.  This non-universality poses
no problem for modeling GRB or blazar high-energy power-law indices,
though observations generally constrain the parameter space to
subluminal or highly-turbulent and modestly superluminal shocks not far
from the Bohm diffusion limit.  Diffusive acceleration at
ultra-relativistic shocks requires \teq{\delta B/B\sim 1} in order to
generate sufficiently low \teq{\sigma} to mesh with observed source
photon spectra.  It is unclear whether conditions in bursts and blazars
can support such levels of turbulence at shocks embedded in their
relativistic outflows, though it should be noted that large field
fluctuations naturally emerge in PIC simulations of Weibel
instability-driven relativistic shocks. It is anticipated that the rich
prospects for {\it Fermi} gamma-ray observations of blazars and GRBs in
the next few years will enhance our understanding of particle
acceleration in their environs.

\bibliographystyle{aipproc}

\end{document}